\documentclass[aps,prl,citeautoscript,reprint,superscriptaddress,floatfix,footinbib,showkeys]{revtex4-2}
\usepackage{setspace}
\usepackage[utf8]{inputenc}
\usepackage[colorlinks=true,linkcolor=black,urlcolor=black,filecolor=black,citecolor=black]{hyperref}
\usepackage{color,soul}
\usepackage{amsmath}
\usepackage{amsfonts}
\usepackage{amssymb}
\usepackage{graphicx}
\usepackage{dcolumn,multirow}
\usepackage{bm}
\usepackage{subfigure}
\usepackage{booktabs}
\usepackage{float}

\newcommand{\mr}[1]{\mathrm{#1}}

\begin{document}
\title{Revisiting dissipation-driven phase transition in a Josephson junction}
\author{Diego Subero}\email{suberoda123@gmail.com}
\affiliation{PICO Group, QTF Centre of Excellence, Department of Applied Physics, Aalto University School of Science, P.O. Box 13500, 0076 Aalto, Finland}

\author{Yu-Cheng-Chang}
%\affiliation{PICO Group, QTF Centre of Excellence, Department of Applied Physics, Aalto University School of Science, P.O. Box 13500, 0076 Aalto, Finland}

\author{Miguel Monteiro}
%\affiliation{PICO Group, QTF Centre of Excellence, Department of Applied Physics, Aalto University School of Science, P.O. Box 13500, 0076 Aalto, Finland}

\author{Ze-Yan Chen}
%\affiliation{PICO Group, QTF Centre of Excellence, Department of Applied Physics, Aalto University School of Science, P.O. Box 13500, 0076 Aalto, Finland}

\author{Jukka P. Pekola}
\affiliation{PICO Group, QTF Centre of Excellence, Department of Applied Physics, Aalto University School of Science, P.O. Box 13500, 0076 Aalto, Finland}

%\email{diego.suberorengel@aalto.fi}
\begin{abstract}
Despite extensive experimental and theoretical work over several decades, Schmid-Bulgadaev quantum phase transition \cite{schmid1983diffusion, bulgadaev1984phase} remains a subject of debate. Here we revisit this problem by performing systematic experiments on low-frequency current-voltage characteristics of Josephson junctions over a wide range of parameters. The experiments are conducted in a true resistive environment formed by a metallic on-chip resistor located near the junction. Over the parameter range of the experiment, we find that the transition occurs when the resistance crosses the quantum value $h/(4e^2)\simeq 6.5$ k$\Omega$ for Cooper pairs, as originally predicted \cite{schmid1983diffusion, bulgadaev1984phase}. The temperature $T$ of the experiment is naturally non-zero, but our basic theoretical modeling corroborates that the observations under these conditions can serve as the basis for the conclusions made, in particular, the crossover resistance from superconducting to insulating regime is the same as that at $T=0$.
\end{abstract}

\maketitle

%\section{Introduction}

Dissipative effects in quantum systems have captivated researchers for many years due to their profound impact on the properties and behavior of these systems. One of the intriguing phenomena is the phase transition induced by the interaction between the quantum system and its surrounding environment. A notable example is the resistively shunted Josephson junction (RSJJ), which undergoes a superconductor-insulator transition that is influenced by the strength of dissipation. \textcolor{black}{Pioneering works by Schmid \cite{schmid1983diffusion} and Bulgadaev \cite{bulgadaev1984phase} revealed that, in the zero-temperature limit, the Josephson junction JJ would become insulating if the zero-frequency environmental impedance ($\textit{R}_\mr{e} = \operatorname{Re} [Z(\omega\rightarrow 0)]$) exceeds the superconducting resistance quantum $\textit{R}_\mr{Q} = \textit{h}/4\textit{e}^2 \approx 6.5$ k$\Omega$, regardless of the strength of the Josephson coupling $E_\mr{J}$.}

Forty years later, this transition continues generating significant attention both experimentally \cite{subero2023bolometric, murani2020absence, kuzmin2025observation} and theoretically \cite{ Houzet_2020, Burshtein_2021, Morel_2021, houzet2023microwave, Masuki2022, Sepulcre2022, MasukiReply2022, burshtein2023inelastic, leger2023revealing, yokota2023functional, R.Daviet_2023,yeyati_2024, NicolasParis_2025,Giacomelli_2024,giacomelli2025exact,Riwar2024, Glazman2024, kurilovich2025quantum}. Although dc charge transport experiments have mostly supported this transition \cite{yagi1997phase, penttila1999superconductor,penttila2001experiments, Perti2020, subero2023bolometric}, recent microwave experiments have not verified the insulating phase, questioning the validity of the Schmid-Bulgadaev (SB) transition \cite{murani2020absence, JoyezReply}. Our recent bolometric heat transport experiment \cite{subero2023bolometric} \textcolor{black}{and the theoretical model based on the nonequilibrium Green's functions \cite{yeyati_2024}} revealed that the Josephson coupling survives well beyond the predicted Schmid-Bulgadaev (SB) phase transition. Conversely, a reflectometry experiment on a Josephson junction coupled to a long section of a transmission line has provided evidence of the transition by monitoring the line's internal dynamics \cite{kuzmin2025observation}, which is influenced by Josephson dynamics. The previous experimental finding has been supported by theoretical calculations involving finite-frequency response \cite{houzet2023microwave,burshtein2023inelastic} and the utilization of exact diagonalization techniques \cite{Giacomelli_2024}. Although finite frequency measurements seemingly provide evidence of the SB phase transition by mimicking the resistor with an array of Josephson junctions or superconducting cavities, it is important to note that these resonators are far from replicating an actual resistor \cite{pekola2024heat, zhong2019violating}: a one-dimensional JJ array acts as a high-impedance environment with well-defined resonances in its absorption spectrum up to the plasma frequency and becomes purely capacitive above it.

In this letter, we revisit the dissipative phase transition of a JJ in an ohmic environment by performing systematic low-frequency lock-in transport measurements. The devices connected to an environment resistance $R_\mr{e}< R_\mr{Q}$ all exhibit a conductance peak at zero voltage bias, signaling the superconducting behavior of the JJ, whereas those with $R_\mr{e}> R_\mr{Q}$ show a pronounced conductance dip, characteristic of the insulating side of a JJ. %, regardless of the coupling value $E_\mr{J}$.
%Even for a JJ connected to a resistor with resistance $R_{\mr e} = 8.8$ k$\Omega$, we observe slight suppression of conductance, indicating insulating-like behavior of the JJ. According to the theory, one might expect a full suppression of the electrical conductance at $R_{\mr e}= R_{\mr Q}$; however, the nonvanishing temperature in the experiment prevents such an abrupt suppression. 
Additionally, we conducted measurements on a superconducting quantum interference device SQUID, acting upon by tunable magnetic field JJ, which allowed us to explore the influence of Josephson coupling on the phase transition. In light of these findings, we construct a finite temperature phase diagram of a JJ revealing a transition near the critical point $R_{\mr Q}/R_{\mr e}\simeq 1$, \textcolor{black}{on the resolution level of our experiment}, that is insensitive to the ratio $E_\mr{J}/E_\mr{C}$, in agreement with the theoretical predictions of Schmid and Bulgadaev \cite{schmid1983diffusion, bulgadaev1984phase}.

\begin{figure}
	\centering	\includegraphics[width=1\columnwidth]{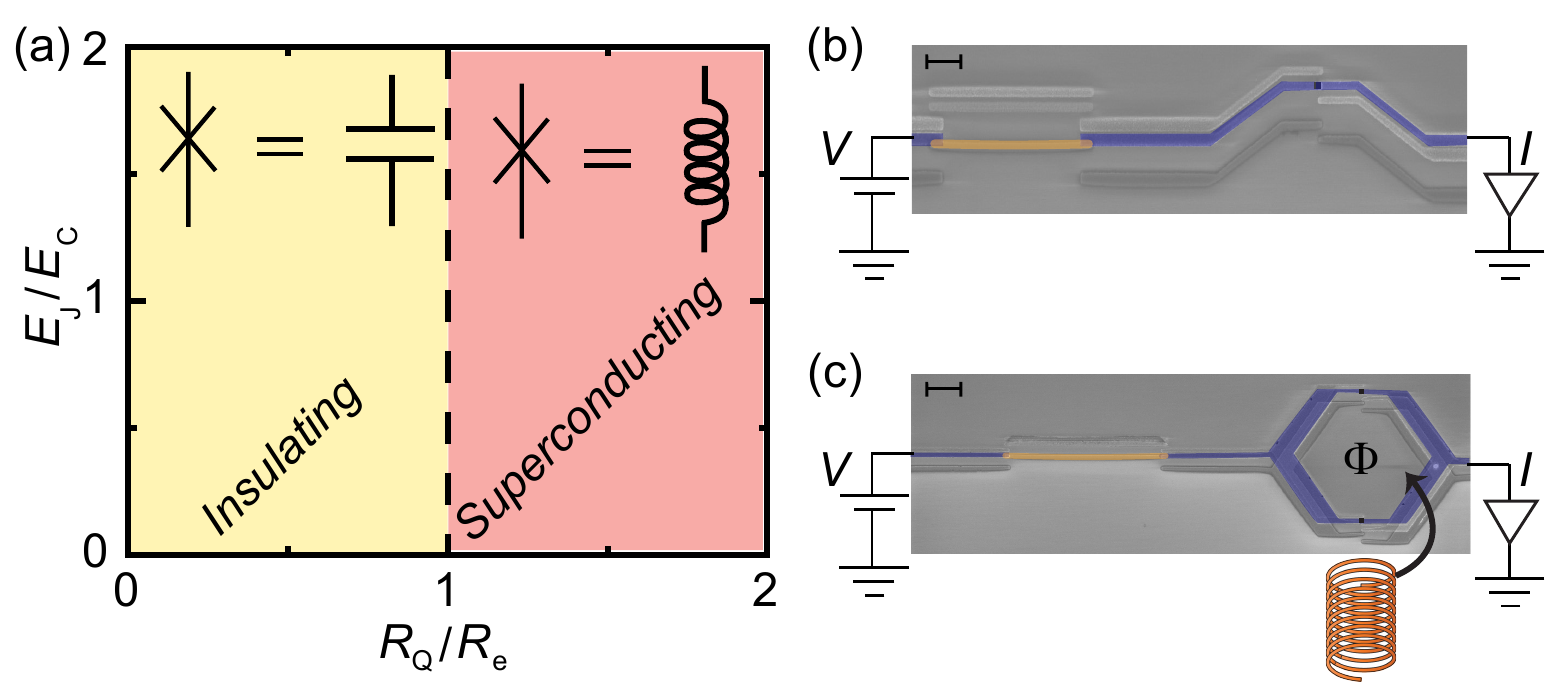} 
	\caption{(a) Schematic illustration of the Schmid–Bulgadaev phase transition in a Josephson junction. False-colored SEM images (scale bar= 2 $\mu$m)  of (b) a single Josephson junction (Al in blue) and (c) a SQUID connected in series to an ohmic environment of resistance $R_{\mr e}$ (Cr in orange). }\label{FigureI}

\end{figure}

Two types of Josephson devices were studied: a single junction with an overlap area of $100 \times 100\ \mu\text{m}^2$ and a SQUID. In both cases, the devices were electrically connected via clean contacts to an on-chip chromium (Cr) resistor that defines the electromagnetic environment [Figs. \ref{FigureI}(b) and \ref{FigureI}(c)]. The samples were fabricated using electron-beam lithography (EBL, Vistec EBPG500+ at 100 kV) combined with a Ge-based hard mask process and triple-angle evaporation \cite{Dolan1977}; see Ref. \cite{subero2023bolometric} for details. The distance between the Josephson element and the resistor was kept to a few micrometers to minimize suppression of environment-induced effects by stray capacitance. The Cr resistance was tuned by varying the resistor length $L$ up to $20\ \mu$m.

For the single junction devices, two fabrication batches were prepared (Table \ref{TableI}). Batch I consisted of three devices on a chip with a tunnel resistance of $22.6\ \text{k}\Omega$, each connected to a Cr resistor of different length, yielding resistances between $2\ \text{k}\Omega$ and $9\ \text{k}\Omega$. Batch II comprised a junction with a tunnel resistance of $\sim 73\ \text{k}\Omega$, connected to Cr resistors of $22.2\ \text{k}\Omega$ and $40.8\ \text{k}\Omega$, respectively. Also for the SQUIDs, two batches were fabricated. Batch I included three devices with varying Josephson junction sizes (Table \ref{table2}), each coupled to an ohmic resistor of approximately $19\ \text{k}\Omega$. Batch II contained two devices connected to resistors of approximately $3.3\ \text{k}\Omega$ and $4.5\ \text{k}\Omega$ (Table \ref{table3}), respectively, enabling exploration of both sides of the Schmid–Bulgadaev transition.

%The devices studied were made of a small single JJ with an overlap area of $100 \times 100 \ \mu\text{m}^2$ or SQUID, both electrically connected via clean contacts to an on-chip chromium (Cr) resistor that defines the electromagnetic environment, as shown in Fig. \ref{FigureI}(b) and \ref{FigureI}(c), respectively. The samples were fabricated by using electron beam lithography (EBL, Vistec EBPG500 + operating at 100 kV) along with a Ge-based hard mask process and triple angle evaporation \cite{Dolan1977}, see Ref. \cite{subero2023bolometric} for more details. A short distance, typically a few microns, is maintained between the Josephson junction and the resistor to prevent the attenuation of the environment-induced effects due to stray capacitance. The Cr resistance was varied by changing the length $L$ of the resistor up to 20 $\mu$m; hence, we fabricated multiple devices on a chip with different lengths, as detailed in Table \ref{TableI}. As for the SQUID, four devices on chip were fabricated. The main difference between them is in the Josephson size, as listed in Table \ref{table2}. These SQUIDs were placed next to an ohmic resistance of about 5.6 k$\Omega$ and $22$ k$\Omega$, allowing us to study both sides of the SB transition. 

\begin{table}
\caption{Main parameters of the measured single junction samples. The junction resistance  $R_{\mr J}$ and environmental resistance $R_{\mr e}$ were determined as detailed in the text. Josephson energy $E_{\mr J}$  is calculated using the Ambegaokar-Baratoff formula, and the charging energy $E_{\mr C}$ was measured from the voltage offset $eV_{\mr {b}}= E_{\mr {C}} \approx 2.74$ K.}
\label{TableI}
\centering
\begin{tabular}{@{}cclccc@{}}
\toprule
Batch & device & \multicolumn{1}{c}{$R_{\mathrm{J}}$ (k$\Omega$)} & $E_{\mathrm{J}}$ (K) & $R_{\mathrm{e}}$ (k$\Omega$) & $L$ ($\mu$m) \\
\midrule
\multirow{3}{*}{I}  & 1 & \multirow{3}{*}{22.6} & \multirow{3}{*}{0.330} & 2.43 & 1  \\
                    & 2 &                       &                        & 4.00 & 2  \\
                    & 3 &                       &                        & 8.80 & 4  \\
\multirow{2}{*}{II} & 4 & \multirow{2}{*}{73.3} & \multirow{2}{*}{0.102} & 22.2 & 10 \\
                    & 5 &                       &                        & 40.8 & 20 \\
\bottomrule
\end{tabular}
\end{table}

\begin{figure*}[ht!]
 \includegraphics[width=\textwidth]{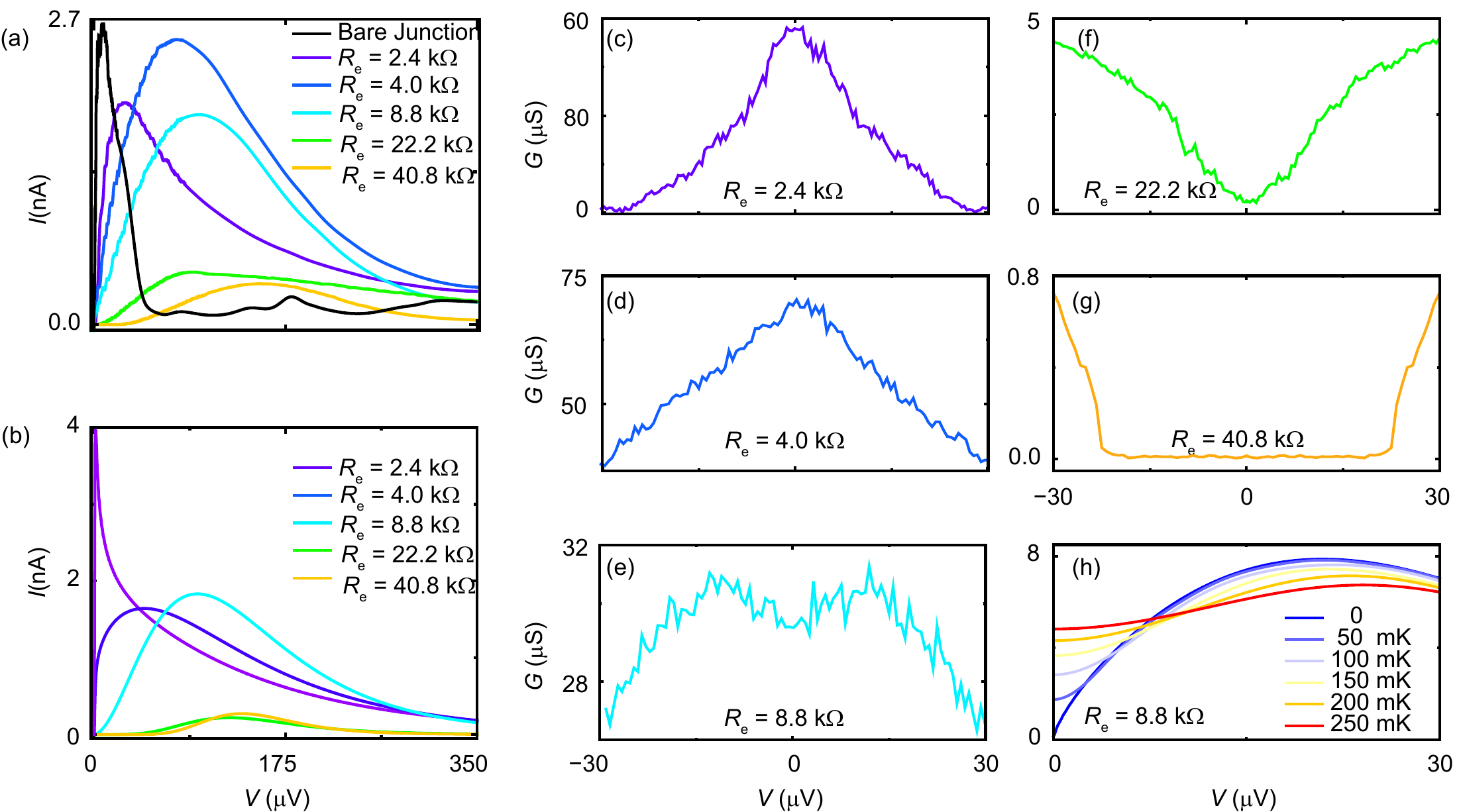}
 	\caption{(a) $IV$-characteristic of a JJ connected to various normal resistors with varying $R_\mr{e}$ at the base temperature of the cryostat (8 mK). The solid black line represents the $IV$ of the bare Josephson junction for batch I. (b) Theoretical results using $P(E)$-theory in an $RC-$ environment. (c), (d), (e), (f), (g).- Differential conductance $G\equiv dI/dV$ near zero voltage bias. (h) Effect of the temperature on the conductance $G\equiv dI/dV$ calculated within the $P(E)$-theory for a Josephson junction coupled to an environment with resistance $R_{\mr e}= 8.8$ k$\Omega$. \textcolor{black}{As expected, the conductance drops to zero at absolute zero temperature, but remains nonvanishing at finite temperatures.} Thus the weak drop at $V= 0$  observed in panel (e) can be attributed to the influence of temperature.}\label{Fig2}
\end{figure*}

\begin{figure*}[ht!]
    \centering
    \includegraphics[width=\textwidth]{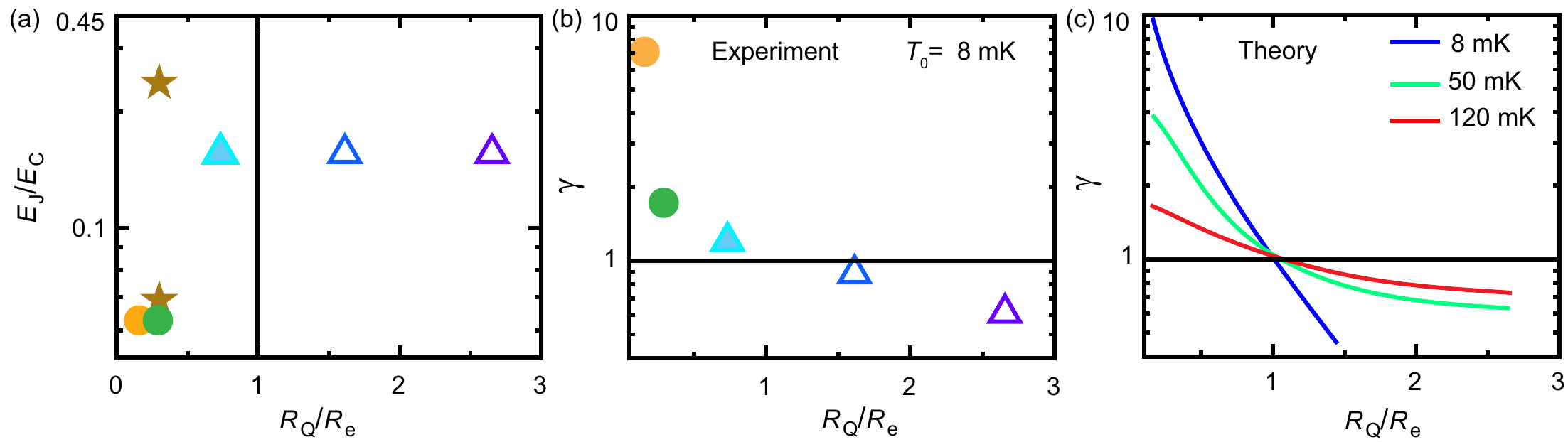}
    \caption{(a) SB phase transition of a Josephson junction. Triangles and circles correspond to data from this work for sample batches I and II, respectively, while stars indicate our prior results from Ref. \cite{subero2023bolometric}. Open (solid) symbols denote superconducting (insulating) behavior, respectively. (b) Power-law scaling (exponent $\gamma$) extracted from the \textit{IV} characteristic shown in Fig. \ref{Fig2}(a) in the very low bias voltage regime $eV<< R_\mr{Q}E_\mr{C}/R_\mr{e}$ as a function of $R_\mr{Q}/R_\mr{e}$. (c) Theoretical predictions for $\gamma$ based on the $P(E)-$theory at various temperatures.}
    \label{phase_diagram}
\end{figure*}

The quasiparticle tunnel resistance $R_\mr{J}$ was independently determined from the $IV$ characteristic of a bare Josephson junction on-chip with a nominally the same size as that used in the composed device, Fig. \ref{FigureI}(b). Moreover, to ensure the accuracy of our measurements, we have also subtracted $R_\mr{J}$ from the total resistance obtained by fitting the $IV$ curve of the composed device at high voltage bias as \textcolor{black}{the sum of the two resistances in series} $R_\mr{T}= R_\mr{J}+ R_\mr{e}$, leading to consistent results. The resistance of the Cr-strip $R_{\mr e}$ was obtained the same way as $R_{\mr J}$, with values listed in Table \ref{TableI}. The Josephson energy $E_{\mr J}$ was calculated using the Ambegaokar-Baratoff relation as $E_{\mr J} = \pi\hbar\Delta/4e^2R_{\mr J}$, with $\Delta\approx 200$ $\mu e$V  being the aluminum superconducting energy gap. \textcolor{black}{The charging energy $E_{\mr {C}}= e^2/2C$ was obtained experimentally from the offset voltage bias $eV_{\mr {b}}= E_{\mr {C}}$ observed at large voltage bias in the $IV$ curve of a bare JJ on-chip in the normal state \cite{ingold1992charge}.} %(see Supplemental information).

\begin{table}
\caption{Experimental and fitting parameters of the 3 SQUID devices connected to an $R_{\mr e}= 19.2\  \text{k}\Omega$ resistor. The SQUID loop area is nominally the same ($6\times6 \ \mu$m$^2$) for the three devices. The difference lies in the size of the Josephson junction. The fitting parameters were obtained using the $P(E)-$theory for normal junctions \cite{ingold1992charge}.}\label{Table_II}
\label{table2}
\begin{tabular}{lllllll}
\hline
Device & JJ area $(\text{nm}^2)$& $R_\mathrm{J}$(k$\Omega$) & $C^\mathrm{Fit}$ (fF)& $E_\mathrm{J}$ (K) & $E_\mathrm{C}$ (K)\\ \hline
(a)   &  $180\times 50$   &20.49                         & 0.80       & 0.375         & 1.20   \\
(b)   &$150\times 50$ &51.80                      & 0.42       & 0.145       & 2.21   \\
(c)     & $100\times 50$ &62.87                       & 0.40       & 0.119       & 2.30   \\ \hline
\end{tabular}
\end{table}

Figure \ref{Fig2}(a) shows a close-up view of the $IV$ curves of a JJ coupled to an ohmic resistor with different resistance values, measured at 7 mK. As can be seen, the current peak \textcolor{black}{at $V= 0$ characteristic of the Josephson current} is shifted to a finite voltage as $R_\mr{e}$ increases, as compared with that \textcolor{black}{in a bare single junction} (solid black line). $R_{\mr e}$ tends to gradually suppress the electrical current of the system at low voltages, which is most notable when  $R_{\mr e}> R_{\mr Q}$. This behavior is well-understood within the so-called $P(E)$ theory \cite{ingold1992charge} in $RC-$environment, as shown in panel (b) of Fig. \ref{Fig2}: the presence of the environment impedes charge relaxation in Cooper-pair tunneling through the junction, resulting in suppressed conductance. \textcolor{black}{One possible reason for the quantitative discrepancy between the data and the theory is the bias-dependent temperature in the experiment.} This behavior becomes more evident in the differential conductance measurements displayed in Figs. \ref{Fig2} (c), (d), (e), and (f).  For the devices with $R_{\mr e}< R_{\mr Q}$, Figs. \ref{Fig2}(c) and (d), a conductance peak developed at zero voltage bias is observed, indicative of the persistence of the critical current (Josephson effect). In contrast, for the devices with $R_{\mr e} > R_{\mr Q}$  as seen in Figs. \ref{Fig2}(e), (f), and (g), the conductance peak tends to be suppressed, signaling the breakdown of the Josephson coupling and the emergence of the insulating behavior. This behavior arises because the superconducting phase becomes delocalized due to increased voltage noise in the electromagnetic environment.  We relate this behavior to the framework predicted by Schmid \cite{schmid1983diffusion} and Bulgadaev \cite{bulgadaev1984phase}. Particularly in our measurements, the transition starts to be visible for the device with $R_{\mr e}=  8.8$ k$\Omega$, shown in Fig. \ref{Fig2}(e), consistent with the theory.
Nevertheless, for $R_{\mr e} \gtrsim R_{\mr Q}$ one might expect an abrupt drop to zero of the conductance at $V= 0$, but such behavior is only anticipated at $T= 0$; hence, the deviation of conductance from zero observed in Fig. \ref{Fig2}(e) can be attributed to the effect of temperature. This assumption is supported by theoretical results obtained by using the $P(E)$-theory as demonstrated in Fig. \ref{Fig2}(g) and agrees with previous theoretical works in Refs. \cite{herrero2002superconductor,kimura2004temperature}.

\begin{table}
\caption{\textcolor{black}{Parameters of the two SQUID devices with identical Josephson junction area, each connected to Cr resistors of \(R_\mathrm{e} = 3.3~\mathrm{k}\Omega\) and \(4.5~\mathrm{k}\Omega\), respectively. Here, the charging energy \(E_\mr{C}\) is estimated from the geometric capacitance $C^{\square}
= 50$ fF/$\mu$m$^2\times$A, where \(A\) is the junction area.}}
\label{table3}
\begin{tabular}{lllllll}
\hline
JJ area $(\text{nm}^2)$& $R_\mathrm{e}$(k$\Omega$) & $R_\mathrm{J}$(k$\Omega$) & $C^{\square}
 $(fF) & $E_\mathrm{J}$ (K) & $E_\mathrm{C}$ (K)\\ \hline
$60\times 60$    &  3.3  & 12.4                         & 0.18       & 0.60        & \textcolor{black}{5.16}  \\
$60\times 60$    &4.5 &9.6                      & 0.18       &0.78       & \textcolor{black}{5.16} \\ \hline
\end{tabular}
\end{table}

\begin{figure*}[ht!]
    \centering
    \includegraphics[width=\textwidth]{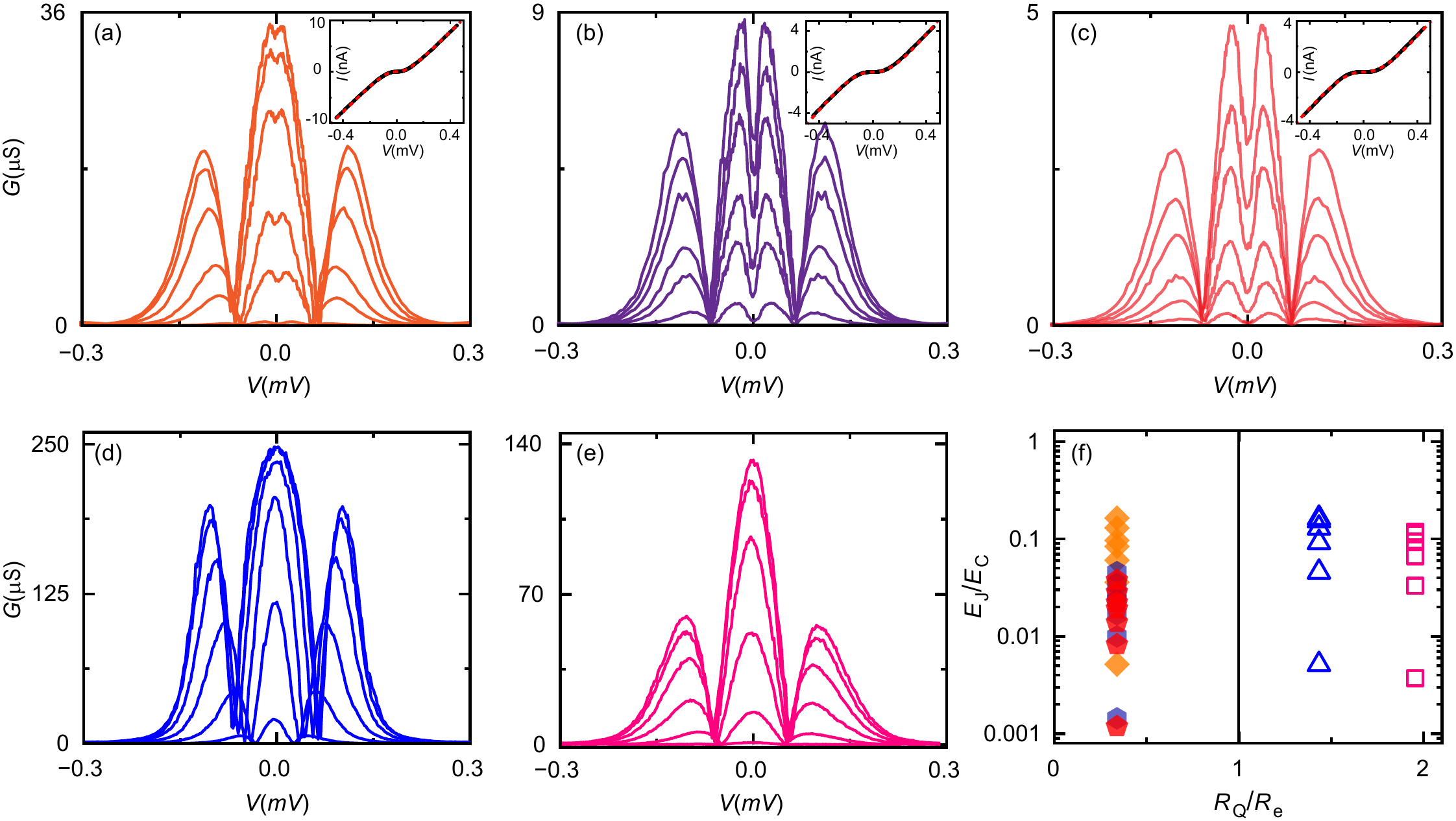}
    \caption{\textcolor{black}{ (a)-(c) Differential conductance of three SQUID devices connected to a Cr resistor with \(R_\mathrm{e} \simeq 19~\mathrm{k}\Omega\) as a function of bias voltage at various normalized magnetic flux values \(\Phi/\Phi_0\)= 0, 0.1, 0.2, 0.3, 0.4, 0.5, from top to bottom. The inset shows the normal-state \(IV\) characteristics at 8 mK, where the solid red line represents the \(P(E)\) theory fit, allowing us to determine the junction capacitance within an \(RC\) environment. 
    (d)-(e) Differential conductance of another two SQUIDs coupled to Cr resistors with \(R_\mathrm{e} \simeq 3.3~\mathrm{k}\Omega\) and \(\simeq 4.5~\mathrm{k}\Omega\), respectively, as a function of bias voltage for different \(\Phi/\Phi_0\).
    (f) Zero-bias conductance of all devices mapped onto the Schmid-Bulgadaev phase diagram where $E_\mr{J}/E_\mr{C}$ is controlled by varying magnetic flux.
    }}
    \label{SQUID_conductance}
\end{figure*}

To map out the SB phase diagram, we have plotted the zero bias conductance character (indicating of superconducting or insulating behavior) for our samples on the coordinate plane with axes $R_{\mr Q}/R_{\mr e}$, $E_{\mr J}/E_{\mr C}$, represented by triangles (batch I) and circles (batch II) in Fig. \ref{phase_diagram}(a). As previously mentioned, the insulating nature of a JJ manifests as conductance suppression at zero bias voltage. The solid line denotes the phase transition by definition. As can be seen, devices 1 and 2 are placed on the superconducting side of the phase diagram (open triangles), while devices 3, 4, and 5 fall on the insulating side, denoted by solid triangles and circles, respectively. The response of the Josephson junction in the presence of a dissipative environment with different resistance values is thus consistent with that predicted by Schmid and Bulgadaev. 

Comparing our phase diagram with the existing experimental phase diagram reported by Yagi \textit{et al.} \cite{yagi1997phase} and Subero \textit{et al.} \cite{subero2023bolometric} (stars) shows that the results in \cite{yagi1997phase} align with theoretical expectations. Yet the two devices with very low $E_{\mr J}/E_{\mr C}= 0.08$ exhibit insulating behavior (solid squares in ref. \cite{yagi1997phase}). This disagreement could be explained by the fact that the Josephson energy was of the order of the temperature at which the experiment was carried out. As a result, the Josephson phase can evolve due to thermal fluctuations, leading to suppression of tunneling by the environmental resistance. Conversely, in Subero \textit{et al.} \cite{subero2023bolometric} charge transport results agree well with the theoretical predictions. 

To complement the above discussion, one might see the phase transition as a change in the slope of the $IV$ curve, following a power law $I\propto V^\gamma$ that holds at $T= 0$ and very low voltage bias $eV<< R_{\mr Q}E_\mr{C}/R_{\mr e}$, with $\gamma= 2R_{\mr e}/R_{\mr Q} -1$, representing the critical exponent associated with the transition. Figure \ref{phase_diagram}(b) displays the dependence of $\gamma$ as a function of $R_{\mr Q}/R_{\mr e}$ from our experiment. As anticipated, the phase transition occurs at $\gamma= 1$, \textit{i.e.} at $R_{\mr e}= R_{\mr Q}$. %Although our $\gamma$ values may not align perfectly with those derived using $P(E)-$theory as shown in Fig. \ref{phase_diagram}c, at least it distinct unambiguously the two phases. This discrepancy can be attributed to the effect of temperature; even though our experiment was conducted at a cryostat temperature of 7 mK, the temperature on the sample remains slightly higher \cite{Giazotto2006}, leading to a deviation from the theoretical prediction.
The influence of temperature on the exponent $\gamma$ based on $P(E)-$theory is shown in Fig. \ref{phase_diagram}(c), with the top-to-bottom lines representing the behavior of $\gamma$ from the lowest to highest temperature. It is worth noting that the crossover point of the transition is unaffected by temperature below 120 mK. %This alignment between experimental and theoretical findings solidifies our understanding of this intriguing phase transition.

\textcolor{black}{ To investigate the dependence of the crossover on the ratio $E_\mr{J}/E_\mr{C}$, we have conducted measurements on a set of SQUID devices with parameters listed in Table \ref{table3}. Figures \ref{SQUID_conductance}(a)-(c) show the differential conductance $G$ as a function of bias voltage for three SQUID devices connected to a Cr-resistance of $R_\mr{e}\simeq 19.2 \ \text{k}\Omega$.  Remarkably, all devices exhibit a pronounced dip in conductance at $V= 0$, regardless of the applied magnetic flux, highlighting the insulating character of the Josephson junction. Here, the junction capacitance was extracted by fitting the normal-state \textit{IV} characteristic with \(P(E)\)-theory \cite{ingold1992charge}, as shown in the inset of figures \ref{SQUID_conductance}(a)-(c), where superconductivity was suppressed using a permanent magnet placed beneath the sample holder.
On the other hand, Figs. \ref{SQUID_conductance}(d)  and (e) display the differential conductance of SQUID devices coupled to a lower resistance, $R_\mathrm{e}= 3.3\ \text{k}\Omega$ and $R_\mathrm{e}= 4.5\ \text{k}\Omega$, respectively. Here, a pronounced peak emerges at zero bias voltage, consistent with a non-vanishing critical current and indicative of coherent Josephson tunneling in the superconducting regime. These observations exemplify the phase–charge duality: in low-impedance environments (superconducting regime), the phase across the junction is well-defined and the charge fluctuates, while in high-impedance environments (insulating regime), the charge becomes localized and the phase undergoes large fluctuations. 
The zero-bias anomaly is represented on a plane $R_\mathrm{Q}/R_\mathrm{e}, \ E_\mathrm{J}/E_\mathrm{C}$, as depicted in Fig. \ref{SQUID_conductance}(f), illustrating the robustness of the insulating (superconducting) behavior across the entire range of $E_\mr{J}/E_\mr{C}$
explored. }

In summary, our study investigates the superconducting-insulating phase transition of a Josephson junction, with a focus on the dc response when coupled to a well-defined dissipative environment. Our findings systematically provide strong evidence that the transition occurs at $R_{\mr e}\approx R_{\mr Q}$, regardless of the strength of the Josephson coupling. Moreover, the current-voltage characteristic exhibits a power-law scaling near zero bias voltage, consistent with a quantum phase transition.
These findings contribute to the ongoing debate over the existence of a dissipative phase transition in Josephson junctions. The apparent discrepancies in the literature likely stem from differences in the frequency regimes in which the experiments are conducted, as highlighted by earlier experimental \cite{murani2020absence,leger2019observation,kuzmin2025observation} and theoretical \cite{houzet2023microwave,burshtein2023inelastic} studies.
%The root of this discrepancy appears to be in the frequency regime in which the experiments are conducted, as elucidated by previous experimental \cite{murani2020absence, leger2019observation,kuzmin2025observation} and theoretical \cite{houzet2023microwave, burshtein2023inelastic} results. 

\textcolor{black}{We acknowledge valuable discussions with A. Levy Yeyati, M. Houzet, and L. Glazman. This work was supported by the Research Council of Finland Centre of Excellence program grants 336810 and grant 349601 (THEPOW) and utilized the OtaNano research infrastructure.}

%\bibliography{Reference}

\begin{thebibliography}{35}%
\makeatletter
\providecommand \@ifxundefined [1]{%
 \@ifx{#1\undefined}
}%
\providecommand \@ifnum [1]{%
 \ifnum #1\expandafter \@firstoftwo
 \else \expandafter \@secondoftwo
 \fi
}%
\providecommand \@ifx [1]{%
 \ifx #1\expandafter \@firstoftwo
 \else \expandafter \@secondoftwo
 \fi
}%
\providecommand \natexlab [1]{#1}%
\providecommand \enquote  [1]{``#1''}%
\providecommand \bibnamefont  [1]{#1}%
\providecommand \bibfnamefont [1]{#1}%
\providecommand \citenamefont [1]{#1}%
\providecommand \href@noop [0]{\@secondoftwo}%
\providecommand \href [0]{\begingroup \@sanitize@url \@href}%
\providecommand \@href[1]{\@@startlink{#1}\@@href}%
\providecommand \@@href[1]{\endgroup#1\@@endlink}%
\providecommand \@sanitize@url [0]{\catcode `\\12\catcode `\$12\catcode
  `\&12\catcode `\#12\catcode `\^12\catcode `\_12\catcode `\%12\relax}%
\providecommand \@@startlink[1]{}%
\providecommand \@@endlink[0]{}%
\providecommand \url  [0]{\begingroup\@sanitize@url \@url }%
\providecommand \@url [1]{\endgroup\@href {#1}{\urlprefix }}%
\providecommand \urlprefix  [0]{URL }%
\providecommand \Eprint [0]{\href }%
\providecommand \doibase [0]{https://doi.org/}%
\providecommand \selectlanguage [0]{\@gobble}%
\providecommand \bibinfo  [0]{\@secondoftwo}%
\providecommand \bibfield  [0]{\@secondoftwo}%
\providecommand \translation [1]{[#1]}%
\providecommand \BibitemOpen [0]{}%
\providecommand \bibitemStop [0]{}%
\providecommand \bibitemNoStop [0]{.\EOS\space}%
\providecommand \EOS [0]{\spacefactor3000\relax}%
\providecommand \BibitemShut  [1]{\csname bibitem#1\endcsname}%
\let\auto@bib@innerbib\@empty
%</preamble>
\bibitem [{\citenamefont {Schmid}(1983)}]{schmid1983diffusion}%
  \BibitemOpen
  \bibfield  {author} {\bibinfo {author} {\bibfnamefont {A.}~\bibnamefont
  {Schmid}},\ }\bibfield  {title} {\bibinfo {title} {Diffusion and localization
  in a dissipative quantum system},\ }\href@noop {} {\bibfield  {journal}
  {\bibinfo  {journal} {Phys. Rev. Lett}\ }\textbf {\bibinfo {volume} {51}},\
  \bibinfo {pages} {1506} (\bibinfo {year} {1983})}\BibitemShut {NoStop}%
\bibitem [{\citenamefont {Bulgadaev}(1984)}]{bulgadaev1984phase}%
  \BibitemOpen
  \bibfield  {author} {\bibinfo {author} {\bibfnamefont {S.}~\bibnamefont
  {Bulgadaev}},\ }\bibfield  {title} {\bibinfo {title} {Phase diagram of a
  dissipative quantum system},\ }\href@noop {} {\bibfield  {journal} {\bibinfo
  {journal} {JETP Lett}\ }\textbf {\bibinfo {volume} {39}},\ \bibinfo {pages}
  {264} (\bibinfo {year} {1984})}\BibitemShut {NoStop}%
\bibitem [{\citenamefont {Subero}\ \emph {et~al.}(2023)\citenamefont {Subero},
  \citenamefont {Maillet}, \citenamefont {Golubev}, \citenamefont {Thomas},
  \citenamefont {Peltonen}, \citenamefont {Karimi}, \citenamefont
  {Mar{\'\i}n-Su{\'a}rez}, \citenamefont {Yeyati}, \citenamefont {S{\'a}nchez},
  \citenamefont {Park} \emph {et~al.}}]{subero2023bolometric}%
  \BibitemOpen
  \bibfield  {author} {\bibinfo {author} {\bibfnamefont {D.}~\bibnamefont
  {Subero}}, \bibinfo {author} {\bibfnamefont {O.}~\bibnamefont {Maillet}},
  \bibinfo {author} {\bibfnamefont {D.~S.}\ \bibnamefont {Golubev}}, \bibinfo
  {author} {\bibfnamefont {G.}~\bibnamefont {Thomas}}, \bibinfo {author}
  {\bibfnamefont {J.~T.}\ \bibnamefont {Peltonen}}, \bibinfo {author}
  {\bibfnamefont {B.}~\bibnamefont {Karimi}}, \bibinfo {author} {\bibfnamefont
  {M.}~\bibnamefont {Mar{\'\i}n-Su{\'a}rez}}, \bibinfo {author} {\bibfnamefont
  {A.~L.}\ \bibnamefont {Yeyati}}, \bibinfo {author} {\bibfnamefont
  {R.}~\bibnamefont {S{\'a}nchez}}, \bibinfo {author} {\bibfnamefont
  {S.}~\bibnamefont {Park}}, \emph {et~al.},\ }\bibfield  {title} {\bibinfo
  {title} {Bolometric detection of {J}osephson inductance in a highly resistive
  environment},\ }\href@noop {} {\bibfield  {journal} {\bibinfo  {journal}
  {Nat. Commun}\ }\textbf {\bibinfo {volume} {14}},\ \bibinfo {pages} {7924}
  (\bibinfo {year} {2023})}\BibitemShut {NoStop}%
\bibitem [{\citenamefont {Murani}\ \emph {et~al.}(2020)\citenamefont {Murani},
  \citenamefont {Bourlet}, \citenamefont {Le~Sueur}, \citenamefont {Portier},
  \citenamefont {Altimiras}, \citenamefont {Esteve}, \citenamefont {Grabert},
  \citenamefont {Stockburger}, \citenamefont {Ankerhold},\ and\ \citenamefont
  {Joyez}}]{murani2020absence}%
  \BibitemOpen
  \bibfield  {author} {\bibinfo {author} {\bibfnamefont {A.}~\bibnamefont
  {Murani}}, \bibinfo {author} {\bibfnamefont {N.}~\bibnamefont {Bourlet}},
  \bibinfo {author} {\bibfnamefont {H.}~\bibnamefont {Le~Sueur}}, \bibinfo
  {author} {\bibfnamefont {F.}~\bibnamefont {Portier}}, \bibinfo {author}
  {\bibfnamefont {C.}~\bibnamefont {Altimiras}}, \bibinfo {author}
  {\bibfnamefont {D.}~\bibnamefont {Esteve}}, \bibinfo {author} {\bibfnamefont
  {H.}~\bibnamefont {Grabert}}, \bibinfo {author} {\bibfnamefont
  {J.}~\bibnamefont {Stockburger}}, \bibinfo {author} {\bibfnamefont
  {J.}~\bibnamefont {Ankerhold}},\ and\ \bibinfo {author} {\bibfnamefont
  {P.}~\bibnamefont {Joyez}},\ }\bibfield  {title} {\bibinfo {title} {Absence
  of a dissipative quantum phase transition in {J}osephson junctions},\
  }\href@noop {} {\bibfield  {journal} {\bibinfo  {journal} {Phys. Rev. X}\
  }\textbf {\bibinfo {volume} {10}},\ \bibinfo {pages} {021003} (\bibinfo
  {year} {2020})}\BibitemShut {NoStop}%
\bibitem [{\citenamefont {Kuzmin}\ \emph {et~al.}(2025)\citenamefont {Kuzmin},
  \citenamefont {Mehta}, \citenamefont {Grabon}, \citenamefont {Mencia},
  \citenamefont {Burshtein}, \citenamefont {Goldstein},\ and\ \citenamefont
  {Manucharyan}}]{kuzmin2025observation}%
  \BibitemOpen
  \bibfield  {author} {\bibinfo {author} {\bibfnamefont {R.}~\bibnamefont
  {Kuzmin}}, \bibinfo {author} {\bibfnamefont {N.}~\bibnamefont {Mehta}},
  \bibinfo {author} {\bibfnamefont {N.}~\bibnamefont {Grabon}}, \bibinfo
  {author} {\bibfnamefont {R.~A.}\ \bibnamefont {Mencia}}, \bibinfo {author}
  {\bibfnamefont {A.}~\bibnamefont {Burshtein}}, \bibinfo {author}
  {\bibfnamefont {M.}~\bibnamefont {Goldstein}},\ and\ \bibinfo {author}
  {\bibfnamefont {V.~E.}\ \bibnamefont {Manucharyan}},\ }\bibfield  {title}
  {\bibinfo {title} {Observation of the schmid--bulgadaev dissipative quantum
  phase transition},\ }\href@noop {} {\bibfield  {journal} {\bibinfo  {journal}
  {Nat. Phys}\ }\textbf {\bibinfo {volume} {21}},\ \bibinfo {pages} {132}
  (\bibinfo {year} {2025})}\BibitemShut {NoStop}%
\bibitem [{\citenamefont {Houzet}\ and\ \citenamefont
  {Glazman}(2020)}]{Houzet_2020}%
  \BibitemOpen
  \bibfield  {author} {\bibinfo {author} {\bibfnamefont {M.}~\bibnamefont
  {Houzet}}\ and\ \bibinfo {author} {\bibfnamefont {L.~I.}\ \bibnamefont
  {Glazman}},\ }\bibfield  {title} {\bibinfo {title} {Critical fluorescence of
  a transmon at the {S}chmid transition},\ }\href
  {https://doi.org/10.1103/PhysRevLett.125.267701} {\bibfield  {journal}
  {\bibinfo  {journal} {Phys. Rev. Lett.}\ }\textbf {\bibinfo {volume} {125}},\
  \bibinfo {pages} {267701} (\bibinfo {year} {2020})}\BibitemShut {NoStop}%
\bibitem [{\citenamefont {Burshtein}\ \emph {et~al.}(2021)\citenamefont
  {Burshtein}, \citenamefont {Kuzmin}, \citenamefont {Manucharyan},\ and\
  \citenamefont {Goldstein}}]{Burshtein_2021}%
  \BibitemOpen
  \bibfield  {author} {\bibinfo {author} {\bibfnamefont {A.}~\bibnamefont
  {Burshtein}}, \bibinfo {author} {\bibfnamefont {R.}~\bibnamefont {Kuzmin}},
  \bibinfo {author} {\bibfnamefont {V.~E.}\ \bibnamefont {Manucharyan}},\ and\
  \bibinfo {author} {\bibfnamefont {M.}~\bibnamefont {Goldstein}},\ }\bibfield
  {title} {\bibinfo {title} {Photon-instanton collider implemented by a
  superconducting circuit},\ }\href
  {https://doi.org/10.1103/PhysRevLett.126.137701} {\bibfield  {journal}
  {\bibinfo  {journal} {Phys. Rev. Lett.}\ }\textbf {\bibinfo {volume} {126}},\
  \bibinfo {pages} {137701} (\bibinfo {year} {2021})}\BibitemShut {NoStop}%
\bibitem [{\citenamefont {Morel}\ and\ \citenamefont
  {Mora}(2021)}]{Morel_2021}%
  \BibitemOpen
  \bibfield  {author} {\bibinfo {author} {\bibfnamefont {T.}~\bibnamefont
  {Morel}}\ and\ \bibinfo {author} {\bibfnamefont {C.}~\bibnamefont {Mora}},\
  }\bibfield  {title} {\bibinfo {title} {Double-periodic josephson junctions in
  a quantum dissipative environment},\ }\href
  {https://doi.org/10.1103/PhysRevB.104.245417} {\bibfield  {journal} {\bibinfo
   {journal} {Phys. Rev. B}\ }\textbf {\bibinfo {volume} {104}},\ \bibinfo
  {pages} {245417} (\bibinfo {year} {2021})}\BibitemShut {NoStop}%
\bibitem [{\citenamefont {Houzet}\ \emph {et~al.}(2024)\citenamefont {Houzet},
  \citenamefont {Yamamoto},\ and\ \citenamefont
  {Glazman}}]{houzet2023microwave}%
  \BibitemOpen
  \bibfield  {author} {\bibinfo {author} {\bibfnamefont {M.}~\bibnamefont
  {Houzet}}, \bibinfo {author} {\bibfnamefont {T.}~\bibnamefont {Yamamoto}},\
  and\ \bibinfo {author} {\bibfnamefont {L.~I.}\ \bibnamefont {Glazman}},\
  }\bibfield  {title} {\bibinfo {title} {Microwave spectroscopy of the schmid
  transition},\ }\href {https://doi.org/10.1103/PhysRevB.109.155431} {\bibfield
   {journal} {\bibinfo  {journal} {Phys. Rev. B}\ }\textbf {\bibinfo {volume}
  {109}},\ \bibinfo {pages} {155431} (\bibinfo {year} {2024})}\BibitemShut
  {NoStop}%
\bibitem [{\citenamefont {Masuki}\ \emph
  {et~al.}(2022{\natexlab{a}})\citenamefont {Masuki}, \citenamefont {Sudo},
  \citenamefont {Oshikawa},\ and\ \citenamefont {Ashida}}]{Masuki2022}%
  \BibitemOpen
  \bibfield  {author} {\bibinfo {author} {\bibfnamefont {K.}~\bibnamefont
  {Masuki}}, \bibinfo {author} {\bibfnamefont {H.}~\bibnamefont {Sudo}},
  \bibinfo {author} {\bibfnamefont {M.}~\bibnamefont {Oshikawa}},\ and\
  \bibinfo {author} {\bibfnamefont {Y.}~\bibnamefont {Ashida}},\ }\bibfield
  {title} {\bibinfo {title} {Absence versus presence of dissipative quantum
  phase transition in {J}osephson junctions},\ }\href
  {https://doi.org/10.1103/PhysRevLett.129.087001} {\bibfield  {journal}
  {\bibinfo  {journal} {Phys. Rev. Lett.}\ }\textbf {\bibinfo {volume} {129}},\
  \bibinfo {pages} {087001} (\bibinfo {year} {2022}{\natexlab{a}})}\BibitemShut
  {NoStop}%
\bibitem [{\citenamefont {Sépulcre}\ \emph {et~al.}(2022)\citenamefont
  {Sépulcre}, \citenamefont {Florens},\ and\ \citenamefont
  {Snyman}}]{Sepulcre2022}%
  \BibitemOpen
  \bibfield  {author} {\bibinfo {author} {\bibfnamefont {T.}~\bibnamefont
  {Sépulcre}}, \bibinfo {author} {\bibfnamefont {S.}~\bibnamefont {Florens}},\
  and\ \bibinfo {author} {\bibfnamefont {I.}~\bibnamefont {Snyman}},\ }\href
  {https://doi.org/10.48550/arxiv.2210.00742} {\bibinfo {title} {Comment on
  ``absence versus presence of dissipative quantum phase transition in
  {J}osephson junctions''}} (\bibinfo {year} {2022})\BibitemShut {NoStop}%
\bibitem [{\citenamefont {Masuki}\ \emph
  {et~al.}(2022{\natexlab{b}})\citenamefont {Masuki}, \citenamefont {Sudo},
  \citenamefont {Oshikawa},\ and\ \citenamefont {Ashida}}]{MasukiReply2022}%
  \BibitemOpen
  \bibfield  {author} {\bibinfo {author} {\bibfnamefont {K.}~\bibnamefont
  {Masuki}}, \bibinfo {author} {\bibfnamefont {H.}~\bibnamefont {Sudo}},
  \bibinfo {author} {\bibfnamefont {M.}~\bibnamefont {Oshikawa}},\ and\
  \bibinfo {author} {\bibfnamefont {Y.}~\bibnamefont {Ashida}},\ }\href
  {https://doi.org/10.48550/ARXIV.2210.10361} {\bibinfo {title} {Reply to
  'comment on "absence versus presence of dissipative quantum phase transition
  in {J}osephson junctions''}} (\bibinfo {year}
  {2022}{\natexlab{b}})\BibitemShut {NoStop}%
\bibitem [{\citenamefont {Burshtein}\ and\ \citenamefont
  {Goldstein}(2023)}]{burshtein2023inelastic}%
  \BibitemOpen
  \bibfield  {author} {\bibinfo {author} {\bibfnamefont {A.}~\bibnamefont
  {Burshtein}}\ and\ \bibinfo {author} {\bibfnamefont {M.}~\bibnamefont
  {Goldstein}},\ }\bibfield  {title} {\bibinfo {title} {Inelastic decay from
  integrability},\ }\href@noop {} {\bibfield  {journal} {\bibinfo  {journal}
  {arXiv:2308.15542}\ } (\bibinfo {year} {2023})}\BibitemShut {NoStop}%
\bibitem [{\citenamefont {L{\'e}ger}\ \emph {et~al.}(2023)\citenamefont
  {L{\'e}ger}, \citenamefont {S{\'e}pulcre}, \citenamefont {Fraudet},
  \citenamefont {Buisson}, \citenamefont {Naud}, \citenamefont
  {Hasch-Guichard}, \citenamefont {Florens}, \citenamefont {Snyman},
  \citenamefont {Basko},\ and\ \citenamefont {Roch}}]{leger2023revealing}%
  \BibitemOpen
  \bibfield  {author} {\bibinfo {author} {\bibfnamefont {S.}~\bibnamefont
  {L{\'e}ger}}, \bibinfo {author} {\bibfnamefont {T.}~\bibnamefont
  {S{\'e}pulcre}}, \bibinfo {author} {\bibfnamefont {D.}~\bibnamefont
  {Fraudet}}, \bibinfo {author} {\bibfnamefont {O.}~\bibnamefont {Buisson}},
  \bibinfo {author} {\bibfnamefont {C.}~\bibnamefont {Naud}}, \bibinfo {author}
  {\bibfnamefont {W.}~\bibnamefont {Hasch-Guichard}}, \bibinfo {author}
  {\bibfnamefont {S.}~\bibnamefont {Florens}}, \bibinfo {author} {\bibfnamefont
  {I.}~\bibnamefont {Snyman}}, \bibinfo {author} {\bibfnamefont {D.~M.}\
  \bibnamefont {Basko}},\ and\ \bibinfo {author} {\bibfnamefont
  {N.}~\bibnamefont {Roch}},\ }\bibfield  {title} {\bibinfo {title} {Revealing
  the finite-frequency response of a bosonic quantum impurity},\ }\href@noop {}
  {\bibfield  {journal} {\bibinfo  {journal} {SciPost Physics}\ }\textbf
  {\bibinfo {volume} {14}},\ \bibinfo {pages} {130} (\bibinfo {year}
  {2023})}\BibitemShut {NoStop}%
\bibitem [{\citenamefont {Yokota}\ \emph {et~al.}(2023)\citenamefont {Yokota},
  \citenamefont {Masuki},\ and\ \citenamefont {Ashida}}]{yokota2023functional}%
  \BibitemOpen
  \bibfield  {author} {\bibinfo {author} {\bibfnamefont {T.}~\bibnamefont
  {Yokota}}, \bibinfo {author} {\bibfnamefont {K.}~\bibnamefont {Masuki}},\
  and\ \bibinfo {author} {\bibfnamefont {Y.}~\bibnamefont {Ashida}},\
  }\bibfield  {title} {\bibinfo {title} {Functional-renormalization-group
  approach to circuit quantum electrodynamics},\ }\href@noop {} {\bibfield
  {journal} {\bibinfo  {journal} {Phys. Rev. A}\ }\textbf {\bibinfo {volume}
  {107}},\ \bibinfo {pages} {043709} (\bibinfo {year} {2023})}\BibitemShut
  {NoStop}%
\bibitem [{\citenamefont {Daviet}\ and\ \citenamefont
  {Dupuis}(2023)}]{R.Daviet_2023}%
  \BibitemOpen
  \bibfield  {author} {\bibinfo {author} {\bibfnamefont {R.}~\bibnamefont
  {Daviet}}\ and\ \bibinfo {author} {\bibfnamefont {N.}~\bibnamefont
  {Dupuis}},\ }\bibfield  {title} {\bibinfo {title} {Nature of the schmid
  transition in a resistively shunted josephson junction},\ }\href
  {https://doi.org/10.1103/PhysRevB.108.184514} {\bibfield  {journal} {\bibinfo
   {journal} {Phys. Rev. B}\ }\textbf {\bibinfo {volume} {108}},\ \bibinfo
  {pages} {184514} (\bibinfo {year} {2023})}\BibitemShut {NoStop}%
\bibitem [{\citenamefont {Yeyati}\ \emph {et~al.}(2024)\citenamefont {Yeyati},
  \citenamefont {Subero}, \citenamefont {Pekola},\ and\ \citenamefont
  {S\'anchez}}]{yeyati_2024}%
  \BibitemOpen
  \bibfield  {author} {\bibinfo {author} {\bibfnamefont {A.~L.}\ \bibnamefont
  {Yeyati}}, \bibinfo {author} {\bibfnamefont {D.}~\bibnamefont {Subero}},
  \bibinfo {author} {\bibfnamefont {J.~P.}\ \bibnamefont {Pekola}},\ and\
  \bibinfo {author} {\bibfnamefont {R.}~\bibnamefont {S\'anchez}},\ }\bibfield
  {title} {\bibinfo {title} {Photonic heat transport through a josephson
  junction in a resistive environment},\ }\href
  {https://doi.org/10.1103/PhysRevB.110.L220502} {\bibfield  {journal}
  {\bibinfo  {journal} {Phys. Rev. B}\ }\textbf {\bibinfo {volume} {110}},\
  \bibinfo {pages} {L220502} (\bibinfo {year} {2024})}\BibitemShut {NoStop}%
\bibitem [{\citenamefont {Paris}\ \emph {et~al.}(2025)\citenamefont {Paris},
  \citenamefont {Giacomelli}, \citenamefont {Daviet}, \citenamefont {Ciuti},
  \citenamefont {Dupuis},\ and\ \citenamefont {Mora}}]{NicolasParis_2025}%
  \BibitemOpen
  \bibfield  {author} {\bibinfo {author} {\bibfnamefont {N.}~\bibnamefont
  {Paris}}, \bibinfo {author} {\bibfnamefont {L.}~\bibnamefont {Giacomelli}},
  \bibinfo {author} {\bibfnamefont {R.}~\bibnamefont {Daviet}}, \bibinfo
  {author} {\bibfnamefont {C.}~\bibnamefont {Ciuti}}, \bibinfo {author}
  {\bibfnamefont {N.}~\bibnamefont {Dupuis}},\ and\ \bibinfo {author}
  {\bibfnamefont {C.}~\bibnamefont {Mora}},\ }\bibfield  {title} {\bibinfo
  {title} {Resilience of the quantum critical line in the schmid transition},\
  }\href {https://doi.org/10.1103/PhysRevB.111.064509} {\bibfield  {journal}
  {\bibinfo  {journal} {Phys. Rev. B}\ }\textbf {\bibinfo {volume} {111}},\
  \bibinfo {pages} {064509} (\bibinfo {year} {2025})}\BibitemShut {NoStop}%
\bibitem [{\citenamefont {Giacomelli}\ and\ \citenamefont
  {Ciuti}(2024)}]{Giacomelli_2024}%
  \BibitemOpen
  \bibfield  {author} {\bibinfo {author} {\bibfnamefont {L.}~\bibnamefont
  {Giacomelli}}\ and\ \bibinfo {author} {\bibfnamefont {C.}~\bibnamefont
  {Ciuti}},\ }\bibfield  {title} {\bibinfo {title} {Emergent quantum phase
  transition of a {J}osephson junction coupled to a high-impedance multimode
  resonator},\ }\href@noop {} {\bibfield  {journal} {\bibinfo  {journal} {Nat.
  Commun}\ }\textbf {\bibinfo {volume} {15}},\ \bibinfo {pages} {5455}
  (\bibinfo {year} {2024})}\BibitemShut {NoStop}%
\bibitem [{\citenamefont {Giacomelli}\ \emph {et~al.}(2025)\citenamefont
  {Giacomelli}, \citenamefont {Devoret},\ and\ \citenamefont
  {Ciuti}}]{giacomelli2025exact}%
  \BibitemOpen
  \bibfield  {author} {\bibinfo {author} {\bibfnamefont {L.}~\bibnamefont
  {Giacomelli}}, \bibinfo {author} {\bibfnamefont {M.~H.}\ \bibnamefont
  {Devoret}},\ and\ \bibinfo {author} {\bibfnamefont {C.}~\bibnamefont
  {Ciuti}},\ }\bibfield  {title} {\bibinfo {title} {Exact duality at low energy
  in a josephson tunnel junction coupled to a transmission line},\ }\href@noop
  {} {\bibfield  {journal} {\bibinfo  {journal} {arXiv preprint
  arXiv:2504.14651}\ } (\bibinfo {year} {2025})}\BibitemShut {NoStop}%
\bibitem [{\citenamefont {Kashuba}\ and\ \citenamefont
  {Riwar}(2024)}]{Riwar2024}%
  \BibitemOpen
  \bibfield  {author} {\bibinfo {author} {\bibfnamefont {O.}~\bibnamefont
  {Kashuba}}\ and\ \bibinfo {author} {\bibfnamefont {R.-P.}\ \bibnamefont
  {Riwar}},\ }\bibfield  {title} {\bibinfo {title} {Limitations of
  caldeira-leggett model for description of phase transitions in
  superconducting circuits},\ }\href
  {https://doi.org/10.1103/PhysRevB.110.184505} {\bibfield  {journal} {\bibinfo
   {journal} {Phys. Rev. B}\ }\textbf {\bibinfo {volume} {110}},\ \bibinfo
  {pages} {184505} (\bibinfo {year} {2024})}\BibitemShut {NoStop}%
\bibitem [{\citenamefont {Remez}\ \emph {et~al.}(2024)\citenamefont {Remez},
  \citenamefont {Kurilovich}, \citenamefont {Rieger},\ and\ \citenamefont
  {Glazman}}]{Glazman2024}%
  \BibitemOpen
  \bibfield  {author} {\bibinfo {author} {\bibfnamefont {B.}~\bibnamefont
  {Remez}}, \bibinfo {author} {\bibfnamefont {V.~D.}\ \bibnamefont
  {Kurilovich}}, \bibinfo {author} {\bibfnamefont {M.}~\bibnamefont {Rieger}},\
  and\ \bibinfo {author} {\bibfnamefont {L.~I.}\ \bibnamefont {Glazman}},\
  }\bibfield  {title} {\bibinfo {title} {Bloch oscillations in a transmon
  embedded in a resonant electromagnetic environment},\ }\href
  {https://doi.org/10.1103/PhysRevB.110.054508} {\bibfield  {journal} {\bibinfo
   {journal} {Phys. Rev. B}\ }\textbf {\bibinfo {volume} {110}},\ \bibinfo
  {pages} {054508} (\bibinfo {year} {2024})}\BibitemShut {NoStop}%
\bibitem [{\citenamefont {Kurilovich}\ \emph {et~al.}(2025)\citenamefont
  {Kurilovich}, \citenamefont {Remez},\ and\ \citenamefont
  {Glazman}}]{kurilovich2025quantum}%
  \BibitemOpen
  \bibfield  {author} {\bibinfo {author} {\bibfnamefont {V.~D.}\ \bibnamefont
  {Kurilovich}}, \bibinfo {author} {\bibfnamefont {B.}~\bibnamefont {Remez}},\
  and\ \bibinfo {author} {\bibfnamefont {L.~I.}\ \bibnamefont {Glazman}},\
  }\bibfield  {title} {\bibinfo {title} {Quantum theory of bloch oscillations
  in a resistively shunted transmon},\ }\href@noop {} {\bibfield  {journal}
  {\bibinfo  {journal} {Nature Communications}\ }\textbf {\bibinfo {volume}
  {16}},\ \bibinfo {pages} {1384} (\bibinfo {year} {2025})}\BibitemShut
  {NoStop}%
\bibitem [{\citenamefont {Yagi}\ \emph {et~al.}(1997)\citenamefont {Yagi},
  \citenamefont {Kobayashi},\ and\ \citenamefont {Ootuka}}]{yagi1997phase}%
  \BibitemOpen
  \bibfield  {author} {\bibinfo {author} {\bibfnamefont {R.}~\bibnamefont
  {Yagi}}, \bibinfo {author} {\bibfnamefont {S.-i.}\ \bibnamefont
  {Kobayashi}},\ and\ \bibinfo {author} {\bibfnamefont {Y.}~\bibnamefont
  {Ootuka}},\ }\bibfield  {title} {\bibinfo {title} {Phase diagram for
  superconductor-insulator transition in single small josephson junctions with
  shunt resistor},\ }\href@noop {} {\bibfield  {journal} {\bibinfo  {journal}
  {Journal of the Physical Society of Japan}\ }\textbf {\bibinfo {volume}
  {66}},\ \bibinfo {pages} {3722} (\bibinfo {year} {1997})}\BibitemShut
  {NoStop}%
\bibitem [{\citenamefont {Penttil{\"a}}\ \emph {et~al.}(1999)\citenamefont
  {Penttil{\"a}}, \citenamefont {Parts}, \citenamefont {Hakonen}, \citenamefont
  {Paalanen},\ and\ \citenamefont {Sonin}}]{penttila1999superconductor}%
  \BibitemOpen
  \bibfield  {author} {\bibinfo {author} {\bibfnamefont {J.}~\bibnamefont
  {Penttil{\"a}}}, \bibinfo {author} {\bibfnamefont {{\"U}.}~\bibnamefont
  {Parts}}, \bibinfo {author} {\bibfnamefont {P.~J.}\ \bibnamefont {Hakonen}},
  \bibinfo {author} {\bibfnamefont {M.}~\bibnamefont {Paalanen}},\ and\
  \bibinfo {author} {\bibfnamefont {E.}~\bibnamefont {Sonin}},\ }\bibfield
  {title} {\bibinfo {title} {“{S}uperconductor-insulator transition” in a
  single {J}osephson junction},\ }\href@noop {} {\bibfield  {journal} {\bibinfo
   {journal} {Phys. Rev. Lett}\ }\textbf {\bibinfo {volume} {82}},\ \bibinfo
  {pages} {1004} (\bibinfo {year} {1999})}\BibitemShut {NoStop}%
\bibitem [{\citenamefont {Penttil{\"a}}\ \emph {et~al.}(2001)\citenamefont
  {Penttil{\"a}}, \citenamefont {Hakonen}, \citenamefont {Sonin},\ and\
  \citenamefont {Paalanen}}]{penttila2001experiments}%
  \BibitemOpen
  \bibfield  {author} {\bibinfo {author} {\bibfnamefont {J.~S.}\ \bibnamefont
  {Penttil{\"a}}}, \bibinfo {author} {\bibfnamefont {P.}~\bibnamefont
  {Hakonen}}, \bibinfo {author} {\bibfnamefont {E.}~\bibnamefont {Sonin}},\
  and\ \bibinfo {author} {\bibfnamefont {M.}~\bibnamefont {Paalanen}},\
  }\bibfield  {title} {\bibinfo {title} {Experiments on dissipative dynamics of
  single josephson junctions},\ }\href@noop {} {\bibfield  {journal} {\bibinfo
  {journal} {Journal of low temperature physics}\ }\textbf {\bibinfo {volume}
  {125}},\ \bibinfo {pages} {89} (\bibinfo {year} {2001})}\BibitemShut
  {NoStop}%
\bibitem [{\citenamefont {Hakonen}\ and\ \citenamefont
  {Sonin}(2021)}]{Perti2020}%
  \BibitemOpen
  \bibfield  {author} {\bibinfo {author} {\bibfnamefont {P.~J.}\ \bibnamefont
  {Hakonen}}\ and\ \bibinfo {author} {\bibfnamefont {E.~B.}\ \bibnamefont
  {Sonin}},\ }\bibfield  {title} {\bibinfo {title} {Comment on ``absence of a
  dissipative quantum phase transition in {J}osephson junctions''},\ }\href
  {https://doi.org/10.1103/PhysRevX.11.018001} {\bibfield  {journal} {\bibinfo
  {journal} {Phys. Rev. X}\ }\textbf {\bibinfo {volume} {11}},\ \bibinfo
  {pages} {018001} (\bibinfo {year} {2021})}\BibitemShut {NoStop}%
\bibitem [{\citenamefont {Murani}\ \emph {et~al.}(2021)\citenamefont {Murani},
  \citenamefont {Bourlet}, \citenamefont {le~Sueur}, \citenamefont {Portier},
  \citenamefont {Altimiras}, \citenamefont {Esteve}, \citenamefont {Grabert},
  \citenamefont {Stockburger}, \citenamefont {Ankerhold},\ and\ \citenamefont
  {Joyez}}]{JoyezReply}%
  \BibitemOpen
  \bibfield  {author} {\bibinfo {author} {\bibfnamefont {A.}~\bibnamefont
  {Murani}}, \bibinfo {author} {\bibfnamefont {N.}~\bibnamefont {Bourlet}},
  \bibinfo {author} {\bibfnamefont {H.}~\bibnamefont {le~Sueur}}, \bibinfo
  {author} {\bibfnamefont {F.}~\bibnamefont {Portier}}, \bibinfo {author}
  {\bibfnamefont {C.}~\bibnamefont {Altimiras}}, \bibinfo {author}
  {\bibfnamefont {D.}~\bibnamefont {Esteve}}, \bibinfo {author} {\bibfnamefont
  {H.}~\bibnamefont {Grabert}}, \bibinfo {author} {\bibfnamefont
  {J.}~\bibnamefont {Stockburger}}, \bibinfo {author} {\bibfnamefont
  {J.}~\bibnamefont {Ankerhold}},\ and\ \bibinfo {author} {\bibfnamefont
  {P.}~\bibnamefont {Joyez}},\ }\bibfield  {title} {\bibinfo {title} {Reply to
  ``comment on `absence of a dissipative quantum phase transition in
  {J}osephson junctions'''},\ }\href
  {https://doi.org/10.1103/PhysRevX.11.018002} {\bibfield  {journal} {\bibinfo
  {journal} {Phys. Rev. X}\ }\textbf {\bibinfo {volume} {11}},\ \bibinfo
  {pages} {018002} (\bibinfo {year} {2021})}\BibitemShut {NoStop}%
\bibitem [{\citenamefont {Pekola}\ and\ \citenamefont
  {Karimi}(2024)}]{pekola2024heat}%
  \BibitemOpen
  \bibfield  {author} {\bibinfo {author} {\bibfnamefont {J.~P.}\ \bibnamefont
  {Pekola}}\ and\ \bibinfo {author} {\bibfnamefont {B.}~\bibnamefont
  {Karimi}},\ }\bibfield  {title} {\bibinfo {title} {Heat bath in a quantum
  circuit},\ }\href@noop {} {\bibfield  {journal} {\bibinfo  {journal}
  {Entropy}\ }\textbf {\bibinfo {volume} {26}},\ \bibinfo {pages} {429}
  (\bibinfo {year} {2024})}\BibitemShut {NoStop}%
\bibitem [{\citenamefont {Zhong}\ \emph {et~al.}(2019)\citenamefont {Zhong},
  \citenamefont {Chang}, \citenamefont {Satzinger}, \citenamefont {Chou},
  \citenamefont {Bienfait}, \citenamefont {Conner}, \citenamefont {Dumur},
  \citenamefont {Grebel}, \citenamefont {Peairs}, \citenamefont {Povey} \emph
  {et~al.}}]{zhong2019violating}%
  \BibitemOpen
  \bibfield  {author} {\bibinfo {author} {\bibfnamefont {Y.~P.}\ \bibnamefont
  {Zhong}}, \bibinfo {author} {\bibfnamefont {H.-S.}\ \bibnamefont {Chang}},
  \bibinfo {author} {\bibfnamefont {K.~J.}\ \bibnamefont {Satzinger}}, \bibinfo
  {author} {\bibfnamefont {M.-H.}\ \bibnamefont {Chou}}, \bibinfo {author}
  {\bibfnamefont {A.}~\bibnamefont {Bienfait}}, \bibinfo {author}
  {\bibfnamefont {C.}~\bibnamefont {Conner}}, \bibinfo {author} {\bibfnamefont
  {{\'E}.}~\bibnamefont {Dumur}}, \bibinfo {author} {\bibfnamefont
  {J.}~\bibnamefont {Grebel}}, \bibinfo {author} {\bibfnamefont {G.~A.}\
  \bibnamefont {Peairs}}, \bibinfo {author} {\bibfnamefont {R.~G.}\
  \bibnamefont {Povey}}, \emph {et~al.},\ }\bibfield  {title} {\bibinfo {title}
  {Violating bell’s inequality with remotely connected superconducting
  qubits},\ }\href@noop {} {\bibfield  {journal} {\bibinfo  {journal} {Nat.
  Phys}\ }\textbf {\bibinfo {volume} {15}},\ \bibinfo {pages} {741} (\bibinfo
  {year} {2019})}\BibitemShut {NoStop}%
\bibitem [{\citenamefont {Dolan}(1977)}]{Dolan1977}%
  \BibitemOpen
  \bibfield  {author} {\bibinfo {author} {\bibfnamefont {G.~J.}\ \bibnamefont
  {Dolan}},\ }\bibfield  {title} {\bibinfo {title} {{Offset masks for lift-off
  photoprocessing}},\ }\href@noop {} {\bibfield  {journal} {\bibinfo  {journal}
  {Appl. Phys. Lett}\ }\textbf {\bibinfo {volume} {31}},\ \bibinfo {pages}
  {337} (\bibinfo {year} {1977})}\BibitemShut {NoStop}%
\bibitem [{\citenamefont {Ingold}\ and\ \citenamefont
  {Nazarov}(1992)}]{ingold1992charge}%
  \BibitemOpen
  \bibfield  {author} {\bibinfo {author} {\bibfnamefont {G.-L.}\ \bibnamefont
  {Ingold}}\ and\ \bibinfo {author} {\bibfnamefont {Y.~V.}\ \bibnamefont
  {Nazarov}},\ }\bibfield  {title} {\bibinfo {title} {Charge tunneling rates in
  ultrasmall junctions},\ }in\ \href@noop {} {\emph {\bibinfo {booktitle}
  {Single charge tunneling}}}\ (\bibinfo  {publisher} {Springer},\ \bibinfo
  {year} {1992})\ pp.\ \bibinfo {pages} {21--107}\BibitemShut {NoStop}%
\bibitem [{\citenamefont {Herrero}\ and\ \citenamefont
  {Zaikin}(2002)}]{herrero2002superconductor}%
  \BibitemOpen
  \bibfield  {author} {\bibinfo {author} {\bibfnamefont {C.~P.}\ \bibnamefont
  {Herrero}}\ and\ \bibinfo {author} {\bibfnamefont {A.~D.}\ \bibnamefont
  {Zaikin}},\ }\bibfield  {title} {\bibinfo {title} {Superconductor-insulator
  quantum phase transition in a single {J}osephson junction},\ }\href@noop {}
  {\bibfield  {journal} {\bibinfo  {journal} {Phys. Rev. B}\ }\textbf {\bibinfo
  {volume} {65}},\ \bibinfo {pages} {104516} (\bibinfo {year}
  {2002})}\BibitemShut {NoStop}%
\bibitem [{\citenamefont {Kimura}\ and\ \citenamefont
  {Kato}(2004)}]{kimura2004temperature}%
  \BibitemOpen
  \bibfield  {author} {\bibinfo {author} {\bibfnamefont {N.}~\bibnamefont
  {Kimura}}\ and\ \bibinfo {author} {\bibfnamefont {T.}~\bibnamefont {Kato}},\
  }\bibfield  {title} {\bibinfo {title} {Temperature dependence of zero-bias
  resistances of a single resistance-shunted {J}osephson junction},\
  }\href@noop {} {\bibfield  {journal} {\bibinfo  {journal} {Phys. Rev. B}\
  }\textbf {\bibinfo {volume} {69}},\ \bibinfo {pages} {012504} (\bibinfo
  {year} {2004})}\BibitemShut {NoStop}%
\bibitem [{\citenamefont {L{\'e}ger}\ \emph {et~al.}(2019)\citenamefont
  {L{\'e}ger}, \citenamefont {Puertas-Mart{\'\i}nez}, \citenamefont
  {Bharadwaj}, \citenamefont {Dassonneville}, \citenamefont {Delaforce},
  \citenamefont {Foroughi}, \citenamefont {Milchakov}, \citenamefont {Planat},
  \citenamefont {Buisson}, \citenamefont {Naud} \emph
  {et~al.}}]{leger2019observation}%
  \BibitemOpen
  \bibfield  {author} {\bibinfo {author} {\bibfnamefont {S.}~\bibnamefont
  {L{\'e}ger}}, \bibinfo {author} {\bibfnamefont {J.}~\bibnamefont
  {Puertas-Mart{\'\i}nez}}, \bibinfo {author} {\bibfnamefont {K.}~\bibnamefont
  {Bharadwaj}}, \bibinfo {author} {\bibfnamefont {R.}~\bibnamefont
  {Dassonneville}}, \bibinfo {author} {\bibfnamefont {J.}~\bibnamefont
  {Delaforce}}, \bibinfo {author} {\bibfnamefont {F.}~\bibnamefont {Foroughi}},
  \bibinfo {author} {\bibfnamefont {V.}~\bibnamefont {Milchakov}}, \bibinfo
  {author} {\bibfnamefont {L.}~\bibnamefont {Planat}}, \bibinfo {author}
  {\bibfnamefont {O.}~\bibnamefont {Buisson}}, \bibinfo {author} {\bibfnamefont
  {C.}~\bibnamefont {Naud}}, \emph {et~al.},\ }\bibfield  {title} {\bibinfo
  {title} {Observation of quantum many-body effects due to zero point
  fluctuations in superconducting circuits},\ }\href@noop {} {\bibfield
  {journal} {\bibinfo  {journal} {Nat. Commun}\ }\textbf {\bibinfo {volume}
  {10}},\ \bibinfo {pages} {5259} (\bibinfo {year} {2019})}\BibitemShut
  {NoStop}%
\end{thebibliography}


%\section{Acknowledgements}
%\textcolor{red}{We thank A. Levy Yeyati, M. Houzet, and L. Glazman for valuable discussions. This work was supported by the Academy of Finland (grant no. 336810) and the ERC under Horizon 2020 (grant agreement no. 742559), and utilized the OtaNano research infrastructure.}

%\section{Author contributions}

%The experiment was conceived by D.S. and J.P.P. and carried out by D.S. with contributions from M. M., Y.  and technical support from J.T.P. Sample fabrication was made by D.S. The theoretical model for heat transport based on the \textit{P(E)} theory was proposed by D.S.G, and the simulations were performed by D.S. The data were analyzed, and the manuscript was written by D.S. with important contributions from all the authors. 

%\section{Competing interests}

%The authors declare no competing interests.
\end{document}